\documentclass[12pt]{article}
\usepackage{a4}
\usepackage{amsmath}
\usepackage{amssymb}
\usepackage[dvips]{graphicx}
\usepackage{subfigure}
\def\beq{\begin{equation}}
\def\eeq{\end{equation}}
\def\beqa{\begin{eqnarray}}
\def\eeqa{\end{eqnarray}}
\def\n{\nonumber \\}

\newcommand {\tr}{{\rm tr}\,}
\newcommand {\Tr}{\mbox{Tr}\,}
\newcommand {\CTr}{{\cal T}r\,}

\def\dag{\dagger}

\begin{document}

\begin{flushright}
{SAGA-HE-240}\\
{KEK-TH-1230}
\end{flushright}
\vskip 0.5 truecm

\begin{center}
{\large{\bf Index theorem in spontaneously symmetry-broken\\
gauge theories on a fuzzy 2-sphere}}
\vskip 1.0cm

{\large Hajime Aoki$^{a}$\footnote{e-mail
 address: haoki@cc.saga-u.ac.jp},
 Yoshiko Hirayama$^{a}$\footnote{e-mail
 address: hirayama@th.phys.saga-u.ac.jp}
 and
 Satoshi Iso$^{b}$\footnote{e-mail
 address: satoshi.iso@kek.jp}
}
\vskip 0.5cm

$^a${\it Department of Physics, Saga University, Saga 840-8502,
Japan  }\\

$^b${\it Institute of Particle and Nuclear Studies, \\
High Energy Accelerator Research Organization (KEK)\\
Tsukuba 305-0801, Japan}

\end{center}

\vskip 1cm
\begin{center}
\begin{bf}
Abstract
\end{bf}
\end{center}
We consider a gauge-Higgs system on a fuzzy 2-sphere
and study the topological structure of gauge configurations,
when the $U(2)$ gauge symmetry is spontaneously broken to
$U(1)\times U(1)$ by the vev of the Higgs field.
The topology is classified
by the index of the Dirac operator satisfying the Ginsparg-Wilson
relation, which turns out to be a noncommutative analog of the
topological charge introduced by 't Hooft.
It can be rewritten as a form
whose commutative limit  becomes
the winding number of the Higgs field.
We also study conditions which assure the validity of the formulation,
and give a generalization of the admissibility condition.
Finally we explicitly calculate the topological charge of 
a one-parameter family of configurations.


\newpage
\setcounter{footnote}{0}
\section{Introduction}
\setcounter{equation}{0}

Matrix models are a promising candidate to 
formulate the superstring theory nonperturbatively \cite{Banks:1996vh,IKKT}, 
where
both spacetime and matter are 
described in terms of matrices, and noncommutative
geometries \cite{Connes} naturally appear \cite{CDS,NCMM,Seiberg:1999vs}. 
One of the important subjects of the matrix model 
is a construction of 
configurations with nontrivial indices in finite noncommutative geometries, 
since compactification of extra dimensions
with nontrivial index can realize 
chiral gauge theory on our spacetime. 
Topologically nontrivial configurations in 
finite noncommutative geometries were constructed 
extensively
\cite{non-trivial_config,non_chi,Valtancoli:2001gx,
Steinacker:2003sd,Karabali:2001te,Carow-Watamura:2004ct}.
A link to relate the topological charge of the background 
to the index of the Dirac operator
is provided by the index theorem \cite{Atiyah:1971rm}.
The index theorem can also be proved in noncommutative ${\mathbb R}^d$
 \cite{Kim:2002qm}.

Extension of the index theorem to finite noncommutative geometry is 
a nontrivial issue due to the doubling problem of the naive Dirac operator.
An analogous problem was solved in the lattice gauge theory
by introducing the Dirac operator which satisfies the
Ginsparg-Wilson (GW) relation \cite{GinspargWilson}.
Its explicit construction was given by
the overlap Dirac operator \cite{Neuberger}
and the perfect action \cite{Hasenfratzindex}.
The exact chiral symmetry \cite{Luscher,Nieder} and 
the index theorem \cite{Hasenfratzindex,Luscher} 
at a finite cutoff can be realized 
due to the GW relation.
These ideas of using the GW relation were also applied 
to the noncommutative geometries.
In ref. \cite{AIN2}, we have provided 
a general prescription to construct 
a GW Dirac operator with coupling to nonvanishing gauge field backgrounds
on general finite noncommutative geometries.
As a concrete example 
we considered the fuzzy 2-sphere \cite{Madore}\footnote{
The GW Dirac operator on the fuzzy 2-sphere for vanishing gauge field 
was given earlier in \cite{balagovi}.
The GW relation was also implemented on the noncommutative 
torus by using the Neuberger's overlap Dirac 
operator \cite{Nishimura:2001dq}.  
In \cite{Iso:2002jc},
this GW Dirac operator was obtained from the general prescription 
\cite{AIN2}.}.
Owing to the GW relation, an index theorem can be proved 
even for finite noncommutative geometries.

We then constructed 't Hooft-Polyakov (TP) monopole configurations
as topologically nontrivial configurations
\cite{Balachandran:2003ay,AIN3}.
We showed that these configurations 
are a noncommutative analogue of the commutative TP monopole 
by explicitly studying the form of the configurations.
We then formulated an index theorem for the TP monopole backgrounds 
by introducing a projection operator
\cite{AIM}.
The topological charge takes the appropriate values
for the TP monopole configurations.
Furthermore, in \cite{AIMN},
we presented a mechanism for
dynamical generation of nontrivial indices,
which may be useful to realize chiral fermion on our spacetime,
by showing that 
the TP monopole configurations with nontrivial topologies
are stabler than the trivial sector
in the Yang-Mills-Chern-Simons matrix model
\cite{Myers:1999ps,IKTW}\footnote{
A related work is given in 
\cite{Steinacker:2007ay}.
The stability of these configurations was
also studied in 
\cite{IKTW,Bal:2001cs,Valtancoli:2002rx,
Imai:2003vr,Azuma:2004zq,
Castro-Villarreal:2004vh,Panero:2006bx}. }.

The index theorem can be extended to general 
configurations which do not obey the equations of motion,
by modifying the chirality operators and the GW Dirac operator
\cite{AIM}.
The topological charge has an appropriate commutative limit, 
introduced by 't Hooft.
Since this formulation is applicable to general configurations
where the $U(2)$ gauge symmetry is  broken down to 
$U(1) \times U(1)$ through the Higgs mechanism, 
the configuration space of gauge fields can be classified 
into the topological sectors. 
Then, all of the topological sectors are defined from a single theory,
while defining the projective module in the noncommutative theories
could provide only a single topological sector\footnote{
A related work is given in \cite{Steinacker:2003sd}.}.

In this paper, we study the topological structure of spontaneously 
symmetry-broken gauge theory on the fuzzy 2-sphere in more detail.
We discuss conditions under which this general formulation is valid. 
This gives a generalization of the admissibility condition,
which was developed in the lattice gauge theory.
We also study some topological properties of the topological charge,
and in particular, we show that the topological charge
is rewritten as a form whose commutative limit becomes 
the winding number of the scalar field. 
Furthermore, as a concrete example, we evaluate the topological charge
and a form of the GW Dirac operator for some explicit 
configurations. 
The results agree with the corresponding commutative cases
if the configurations satisfy the admissibility condition.
We further extend the configurations to the non-admissible regions.

In section \ref{sec:ITforTP}, 
we review the formulation of the index theorem 
for the TP monopole backgrounds.
In section \ref{sec:Form-gen}, 
we study the index theorem for general configurations. 
Some detailed calculations for taking the commutative limits
are sent to appendices \ref{sec:U1} and 
\ref{sec:comlimtopcharge}.
In section \ref{sec:For-L+htau}, 
we investigate the explicit configurations.
Analyses on the zero-modes are given in appendix \ref{sec:anachi}.
Section \ref{conclusion} is devoted to conclusions and discussions.


\section{Formulation for the TP monopole background}
\label{sec:ITforTP}
\setcounter{equation}{0}

In this section, 
we summarize our previous results on the Ginsparg-Wilson (GW) Dirac operator
and the index theorem for the 't Hooft-Polyakov (TP) monopole backgrounds.

\subsection{Dirac operators on fuzzy 2-sphere}

Noncommutative coordinates of the
fuzzy 2-sphere are described by
$x_i =\alpha L_i$, 
where $\alpha$ is the noncommutative parameter,
and $L_i$'s are $n$-dimensional irreducible
representation matrices of $SU(2)$ algebra.
Then we have the relation
$
(x_i)^2
= \rho^2 \mbox{\boldmath  $1$}_n 
$,
where 
$\rho=\alpha \sqrt{(n^2-1)/4}$ 
expresses the radius of the fuzzy 2-sphere.
The commutative limit can be taken by $\alpha \to0, n \to \infty$
with $\rho$ fixed.

Any wave functions on the fuzzy 2-sphere are mapped to 
$n \times n$ matrices. 
We can expand them in terms of noncommutative
analogues of the spherical harmonics. 
Derivatives along the Killing vectors of a function $M(\Omega)$ 
on the 2-sphere 
are written as the adjoint operator of $L_i$ on the corresponding 
matrix $\hat{M}$:
\[
{{\cal L}_i}M(\Omega)=-i\epsilon_{ijk}x_j \partial_k M(\Omega)
\ \leftrightarrow \
{\tilde L}_i \hat{M}= [L_i \, , \, \hat{M}] \ .
\]
An integral of a function is given by a trace
of the corresponding matrix:
\[
\int \frac{d \Omega}{4 \pi}M(\Omega)
\ \leftrightarrow \ \frac{1}{n} \Tr[\hat{M}] \ . 
\]

Two types of Dirac operators have been proposed in 
\cite{Carow-Watamura:1996wg} and \cite{Grosse:1994ed}.
$D_{\rm WW}$ in \cite{Carow-Watamura:1996wg}
has exact chiral symmetry but has doublers.
$D_{\rm GKP}$ in \cite{Grosse:1994ed} has no doublers
but breaks chiral symmetry at finite matrix size.
The chiral anomaly 
is correctly reproduced in the commutative 
limit\cite{non_chi,chiral_anomaly,chiral_anomaly2,AIN1}.

The fermionic action of $D_{\rm GKP}$ is given by
\begin{eqnarray}
S_{\rm GKP}&=& \Tr[\bar\Psi D_{\rm GKP} \Psi] \ , \n
D_{\rm GKP}&=& \sigma_i ({\tilde L}_i + \rho a_i ) +1 \ ,
\label{DGKP}
\end{eqnarray}
where $\sigma_i$'s are Pauli matrices. 
The gauge field $a_i$ of $U(k)$ gauge group
and the fermionic field $\Psi$ in the fundamental representation of the gauge group
are expressed by $nk\times nk$ and $nk\times n$ matrices, respectively.  
This action is invariant under the gauge transformation:
\beq
\Psi \rightarrow U \Psi \ , \ \ \
a_i \rightarrow U a_i U^\dag +\frac{1}{\rho} (U L_i U^\dag-L_i) \ ,
\label{gaugeTra}
\eeq
since a combination,
which is called a covariant coordinate,
\begin{equation}
A_i \equiv L_i+\rho a_i
\label{defAi}
\end{equation}
transforms covariantly as
$
A_i \rightarrow U A_i U^\dagger 
$. 

In the commutative limit, 
the Dirac operator (\ref{DGKP}) becomes 
\begin{equation}
D_{\rm GKP} \rightarrow 
D_{\rm com}=\sigma_i ({\cal L}_i + \rho a_i ) +1 \ ,
\end{equation}
which is the ordinary Dirac operator on the commutative 2-sphere. 
The gauge fields $a_i$'s in 3-dimensional space can be 
decomposed into the tangential components on the 
2-sphere $a_i'$ and the normal component $\phi$ as
\begin{eqnarray}
&&\left\{
\begin{array}{lll}
a_i'&=& \epsilon_{ijk}n_j a_k \ , \\
\phi&=&n_i a_i \ ,
\end{array}
\right. \label{decomposeto}\\
&\Leftrightarrow& a_i = -\epsilon_{ijk}n_j a_k' + n_i \phi \ ,
\label{decomposefrom}
\end{eqnarray}
where $n_i = x_i/\rho$ is a unit vector.
The normal component $\phi$ is a scalar field on the 2-sphere,
and fermions are coupled to the scalar field through 
the Yukawa coupling.


\subsection{GW Dirac operator}

In order to discuss the chiral structures, 
a Dirac operator satisfying the GW relation is more suitable. 
Ref.\cite{AIN2} provided a general prescription to
define a GW Dirac operator in arbitrary gauge 
field backgrounds. 
We first define two chirality operators:
\beq
\Gamma = a\left(\sigma_i L_i^R -\frac{1}{2}\right) \ , 
\ \ \
\hat\Gamma = \frac{H}{\sqrt{H^2}} \label{gammaRgammahat} \ ,
\eeq
with
\begin{equation}
H=a\left(\sigma_i A_i +\frac{1}{2}\right) \ , 
\end{equation}
where
$A_i$ is defined in (\ref{defAi}), and
$a=2/n$
is a noncommutative analogue of a lattice-spacing.
The upper index $R$ in $L_i^R$ means that this 
operator acts from the right on matrices.
These chirality operators satisfy
\begin{equation}
(\Gamma)^\dagger=\Gamma \ , \ \ \
(\hat\Gamma)^\dagger=\hat\Gamma \ , \ \ \
(\Gamma)^2=(\hat\Gamma)^2=1 \ .
\end{equation}
In the commutative limit, both $\Gamma$ and $\hat\Gamma$
become the chirality operator on the commutative 2-sphere,
$\gamma = n_i \sigma_i$.

We then define the GW Dirac operator as
\begin{equation}
D_{\rm GW} = -a^{-1}\Gamma (1- \Gamma \hat{\Gamma}) \ .
\label{defDGW}
\end{equation}
By the definition,
the GW relation  
\begin{equation}
\Gamma D_{\rm GW}+D_{\rm GW} \hat{\Gamma}=0 
\label{GWrelation}
\end{equation}
is satisfied, owing to which
the index theorem can be proved.
The action
\begin{equation}
S_{\rm GW}= \Tr [\bar\Psi D_{\rm GW} \Psi]
\end{equation}
is invariant under the gauge transformation
(\ref{gaugeTra}).
In the commutative limit, $D_{\rm GW}$ becomes 
\begin{equation}
D_{\rm GW} \rightarrow 
D'_{\rm com}=\sigma_i ({\cal L}_i + \rho P_{ij} a_j ) +1 \ ,
\label{DGWcom}
\end{equation}
where $P_{ij}=\delta_{ij}-n_i n_j$ 
is the projector to the tangential directions 
on the sphere. 
This operator $D'_{\rm com}$ is 
the Dirac operator without coupling to the scalar field,
which is consistent with the fact that $D_{\rm GW}$
satisfies the GW relation, a modified chiral symmetry.

\subsection{Monopole configurations}
\label{sec:monopole}

As topologically nontrivial configurations
in the $U(2)$ gauge theory on the fuzzy 2-sphere,
the following monopole configurations were constructed 
\cite{Balachandran:2003ay,AIN3}:
\begin{equation}
A_i=
\begin{pmatrix}
 L_i^{(n+m)} & \cr
& L_i^{(n-m)} \cr
\end{pmatrix} 
\label{LnpmLnmm} \ , 
\end{equation}
where $A_i$ is defined in (\ref{defAi}),
and $L_i^{(n\pm m)}$ are $(n\pm m)$ dimensional 
irreducible representations of $SU(2)$ algebra.
The total matrix size is $N=2n$.
The $m=0$ case corresponds to two coincident fuzzy 2-spheres,
whose effective action is given by the $U(2)$ gauge theory
on the fuzzy 2-sphere.
The cases with  general $m$ correspond to two fuzzy 2-spheres 
with different radii.
For $|m| \ll n$,
they correspond
to the monopole configurations with magnetic charge $-|m|$,
where the $U(2)$ gauge symmetry  is spontaneously broken down
to $U(1) \times U(1)$.

For the $m=1$ case,
(\ref{LnpmLnmm}) is unitary equivalent to 
\begin{equation}
U A_i U^\dagger= L_i^{(n)} \otimes \mbox{\boldmath  $1$}_2 +
\mbox{\boldmath  $1$}_{n} \otimes \frac{\tau_i}{2} \ .
\label{AiLtau}
\end{equation}
Comparing with (\ref{defAi}), the gauge field is given by 
\begin{equation}
a_i=\frac{1}{\rho}\mbox{\boldmath  $1$}_{n} \otimes \frac{\tau_i}{2} \ .
\label{monopoleconfig}
\end{equation}
By taking the commutative limit of (\ref{monopoleconfig}), 
and decomposing it into the normal and the tangential components
of the sphere as in (\ref{decomposeto}), 
it becomes 
\beqa
a_i^{\prime a} &=& \frac{1}{\rho} \epsilon_{ija} n_j \ , 
\label{TPmonConf-g} \\
\phi^a &=& \frac{1}{\rho} n_a \ ,
\label{TPmonConf-h}
\eeqa
which is precisely the TP monopole configuration.

The configuration with $m=0$ is the vacuum configuration,
and of course topologically trivial.
A topologically trivial  configuration with a non-vanishing
 expectation value
of the scalar field is given by
\beq
A_{i}=L_{i}+\frac{2}{n}L_{i}\tau_{3} \ .
\label{trivialconfiginU1U1}
\eeq
Its commutative limit becomes
$a^{\prime a}_i=0, \phi^a=\delta_{a3}/\rho$
and the $U(2)$ gauge symmetry is spontaneously broken to 
$U(1) \times U(1)$. 

Monopole harmonics around the configurations (\ref{LnpmLnmm}) are calculated,
and fiber bundles in matrix models are studied in \cite{ISTT}.

\subsection{Index theorem for the monopole backgrounds}
\label{sec:Index}

The index theorem for the TP monopole backgrounds (\ref{LnpmLnmm})
were formulated \cite{AIM}
as\footnote{
This equation can be proved by using the GW relation 
(\ref{GWrelation}) and the fact that
$P^{(n \pm |m|)}$ commutes with $\Gamma$, $\hat\Gamma$ 
and $D_{\rm GW}$.
}:
\begin{equation}
{\rm{index}}(P^{(n \pm |m|)}D_{\rm GW})=\frac{1}{2}
\CTr[P^{(n \pm |m|)}(\Gamma +\hat{\Gamma})] \ , 
\label{indextheorem}
\end{equation}
where ${\cal T}r$ denotes a trace 
over the space of matrices and over the spinor index.
$P^{(n \pm |m|)}$ is the projection operator to pick up
the Hilbert space for
the $n \pm |m|$ dimensional representation in (\ref{LnpmLnmm}).
That is, it picks up 
one of the two fuzzy 2-spheres.
It is written as
\beq
P^{(n \pm |m|)}
= \frac{1}{2}(1\pm T) \ ,
\label{P1phi}
\eeq
with
\begin{eqnarray}
T &=& \frac{2}{n|m|}\left(A_i^2 - \frac{n^2+m^2-1}{4}\right) \label{defT}\\
  &=& \frac{m}{|m|}\begin{pmatrix}
 \mbox{\boldmath  $1$}_{(n+m)} & \cr
& -\mbox{\boldmath  $1$}_{(n-m)} \cr
\end{pmatrix} \ .
\label{1npm1nmp}
\end{eqnarray}

On the other hand, in the representation (\ref{defAi}), 
 (\ref{defT}) becomes 
\begin{equation}
T  =\frac{2}{n|m|}\left(\rho \{ L_i , a_i \}+ \rho^2 a_i^2-\frac{m^2}{4}\right) \ .
\label{Tother}
\end{equation}
In the commutative limit, $T$ becomes
$\frac{2\rho}{|m|} \phi$ when $|m| \ll n$,
where $\phi$ is the scalar field 
defined in (\ref{decomposeto}).
Moreover, it is normalized as $T^2={\bf 1}_{2n}$. 
Therefore, $T$ is the generator for 
the unbroken $U(1)$ gauge group in the TP monopole.
Recall that the TP monopole configuration breaks
the $SU(2)$ gauge symmetry down to $U(1)$. 
Then, the eigenstate of $T$ with  eigenvalue $\pm 1$
corresponds to the fermionic state with $\pm 1/2$
electric charge of the unbroken $U(1)$ gauge group.
Thus, the index in the projected space (\ref{indextheorem}) 
gives the index for each electric charge component.
Without the projection operator, contributions from $+1/2$ and $-1/2$ 
charges cancel the index, and 
this is why we introduced the projection operator.

The right-hand side (rhs) of (\ref{indextheorem}) has the following properties.
Firstly, it takes only integer values 
since both $\Gamma$ and $\hat{\Gamma}$ have a form of sign operator. 
Secondly,
for the TP monopole configurations (\ref{LnpmLnmm}),
it takes appropriate values
\begin{equation}
\frac{1}{2}{\cal T}r[P^{(n \pm |m|)}(\Gamma +\hat{\Gamma})]
=\mp |m| 
\label{TPpmm}
\end{equation}
for $\pm 1/2$ electric charge component.
Finally,
in the commutative limit, we obtain
\beq
\frac{1}{2}{\cal T}r \left[ \frac{1}{2}T (\Gamma +\hat{\Gamma}) \right] 
\to
\frac{\rho^2}{8\pi}\int_{S^2} d\Omega \ \epsilon_{ijk}
n_i \phi'^a F_{jk}^a \ , 
\label{TCunbrokenprocom}
\eeq
where $\phi'^a$ is a scalar field 
normalized as $\sum_a (\phi'^a)^2=1$.
$F_{jk}=F_{jk}^a \tau^a/2$ 
is the field strength defined as
$F_{jk}= \partial_j a_k'-\partial_k a_j'-i[a_j',a_k']$.
Equation (\ref{TCunbrokenprocom}) is 
the magnetic charge for the unbroken $U(1)$ component
in the TP monopole configuration\footnote{
The topological charge should have an additional term as 
the second term in 
(\ref{TCforTP}).
However this term vanishes for the TP monopole configurations.
}.
From (\ref{TPpmm}) and 
$\frac{1}{2}{\cal T}r[P^{(n \pm |m|)}(\Gamma +\hat{\Gamma})]
=\pm\frac{1}{2}{\cal T}r \left[ \frac{1}{2}T(\Gamma +\hat{\Gamma}) \right]$,
we obtain 
\beq
\frac{1}{2}{\cal T}r \left[ \frac{1}{2}T(\Gamma +\hat{\Gamma}) \right]
=-|m|
\label{topcharnegative-m}
\eeq
for the configurations (\ref{LnpmLnmm}).
Note that only negative topological charge can be 
defined in this formulation.

\section{Formulation for the general configurations}
\label{sec:Form-gen}
\setcounter{equation}{0}

\subsection{Index theorem for general configurations }
\label{3.1}
In the previous section, 
we have considered the index theorem (\ref{indextheorem})
for the monopole background configurations (\ref{LnpmLnmm}),
which satisfy the equations of motion. 
We now extend it to general configurations which do not
necessarily obey the equations of motion.
The only assumption in the following is that the 
$U(2)$ gauge symmetry is spontaneously broken to $U(1) \times U(1)$
through the Higgs mechanism, i.e. a nonzero value of 
the scalar field. 

We first generalize the definition of the operator $T$ in (\ref{defT}) to
\beq
T' = \frac{(A_i)^2-\displaystyle{\frac{n^2-1}{4}}}
{\sqrt{\left[ (A_i)^2-\displaystyle{\frac{n^2-1}{4}} \right]^2}} \ .
\label{defTgen}
\eeq
This definition is valid for general configurations $A_i$
unless the denominator has zero-modes.
For the configurations (\ref{LnpmLnmm}),
$T'$ reduces to the previous one (\ref{1npm1nmp}).
Furthermore, it satisfies 
\beq
(T')^\dag =T' \ , \ (T')^2 = 1 \ ,
\eeq
and then its eigenvalue takes $1$ or $-1$. 
As we show in Appendix \ref{sec:U1},
the commutative limit of $T'$ becomes the normalized scalar field as
\beq
T' \to 2 \phi'=2\phi'^a \frac{\tau^a}{2} \ ,
\label{comlimTprime}
\eeq
where we omitted the $U(1)$ part 
in the $U(2) = SU(2) \times U(1)$ gauge group.
Then, the eigenstate of $T'$ with eigenvalue $\pm 1$
corresponds to the fermionic state with the electric charge
$\pm 1/2$ of unbroken $U(1) \subset SU(2)$ gauge group.
We will then call $T'$ an electric charge operator,
neglecting a factor $1/2$.

We next define modified chirality operators as
\beqa
\Gamma' &=& 
\frac{\{T', \Gamma \}}{\sqrt{\{T', \Gamma \}^2}} 
= T' \Gamma \ , 
\label{defgammaprime}\\
\hat\Gamma' &=& \frac{\{T', \hat\Gamma \}}{\sqrt{\{T', \hat\Gamma \}^2}} \ ,
\label{defgammahatprime}
\eeqa
where $\Gamma$ and $\hat\Gamma$ are defined in
(\ref{gammaRgammahat}).
In (\ref{defgammaprime}), we used $[T', \Gamma] = 0 $.
While these chirality operators are weighted by the electric charge 
operator $T'$,
they still satisfy the usual relations: 
\beq
(\Gamma')^\dag = \Gamma' \ , \ \ \
(\hat\Gamma')^\dag =\hat\Gamma' \ , \ \ \
(\Gamma')^2 = (\hat\Gamma')^2 =1 \ .
\eeq

From these chirality operators, 
we  define a modified GW Dirac operator as
\beq
D'_{\rm GW} = -a^{-1} \Gamma' (1 - \Gamma' \hat\Gamma') \ .
\label{defDGWgen}
\eeq
This Dirac operator is also  weighted by the electric 
charge operator $T'$.
As we show in Appendix \ref{sec:comlimtopcharge}, 
in the commutative limit, this Dirac operator becomes
\beq
D'_{\rm GW} \ \to \ \frac{1}{2} \{2\phi' \, , \, D'_{\rm com} \} \ .
\label{Dcom}
\eeq
In particular, in the $\phi'^a(x) = (0,0,1)$ gauge, it becomes
\beq
\tau^3 \left(\sigma_i {\cal L}_i +1 
+ \rho \sigma_i P_{ij} 
\Bigl(a_j^3 \frac{\tau^3}{2} + a_j^0 \frac{\bf 1}{2}\Bigr)
\right) \ ,
\eeq
which is the Dirac operator with coupling to 
the unbroken $U(1) \times U(1)$ gauge fields, $ a_j^3$ and  $a_j^0$. 

From the definition (\ref{defDGWgen}), 
this Dirac operator satisfies the GW relation
\beq
\Gamma' D'_{\rm GW} + D'_{\rm GW} \hat\Gamma' =0 \ .
\eeq
Thus, the index theorem
\beq
\frac{1}{2} {\rm index}(D'_{\rm GW}) = 
\frac{1}{4} {\cal T}r [\Gamma' + \hat\Gamma']
\label{indextheogen}
\eeq
can be proved similarly to the ordinary case.
Since $\Gamma' $ and $\hat\Gamma'$ are weighted by 
the electric charge operator $T'$,
the cancellation of the index by
the contributions from $\pm1/2$ electric charge components
is avoided.
For the configurations (\ref{LnpmLnmm}),
$T'$ commutes with $\hat\Gamma$,
and then we obtain $\hat\Gamma' = T'\hat\Gamma$.
Thus the rhs of (\ref{indextheogen}) reduces to the previous one,
the left-hand side (lhs) of (\ref{TCunbrokenprocom}).
For the configuration (\ref{trivialconfiginU1U1}),
the rhs of (\ref{indextheogen}) gives a vanishing value.

Furthermore,
as we show in Appendix \ref{sec:comlimtopcharge},
for general configurations,
the commutative limit of the rhs in (\ref{indextheogen}) becomes
\beq
\frac{1}{4} {\cal T}r [\Gamma' + \hat\Gamma']
\to
\frac{\rho^2}{8\pi}\int_{S^2} d\Omega \epsilon_{ijk}
n_i \Bigl( \phi'^a F_{jk}^a 
- \epsilon_{abc} \phi'^a (D_j \phi')^b (D_k \phi')^c \Bigr) \ ,
\label{TCforTP}
\eeq
where $F_{jk}=F_{jk}^a \tau^a/2$ 
is the field strength defined as
$F_{jk}= \partial_j a_k'-\partial_k a_j'-i[a_j',a_k']$,
and $D_{j}$ 
is the covariant derivative defined as
$D_{j} =\partial_{j}  -i[a'_{j} , ~~]$.
$a'_j$ is the tangential components of the gauge field
defined in (\ref{decomposeto}).
This is precisely the topological charge in the case where
the $SU(2)$ gauge symmetry is spontaneously broken down to $U(1)$
\cite{'tHooft:1974qc}.
Hence, the index theorem (\ref{indextheogen})
gives a natural generalization to general 
configurations which are not restricted to 
the special configurations such as the TP monopoles.

\subsection{Admissibility condition}
\label{sec:admissibilitycon}

Now that we have the formulation (\ref{indextheogen}) 
where the topological charge can be defined for general configurations,
the gauge configuration space on the fuzzy 2-sphere can be classified 
into the topological sectors. 
For this, we need to exclude the regions in the configuration space
which separate the different topological sectors.
In the lattice gauge theories,
the admissibility condition was introduced to assure 
this condition and
the locality of the overlap Dirac operator
\cite{Luscher:1981zq,Hernandez:1998et,Luscher:1998kn}. 
Here we will study similar conditions
in the formulation (\ref{indextheogen}).

The formulation (\ref{indextheogen}) is valid
 if the denominators of the three operators,
$T'$ defined in (\ref{defTgen}), $\hat\Gamma$ in (\ref{gammaRgammahat}),
 and  
$\hat\Gamma'$ in (\ref{defgammahatprime}), do not have zero-modes.
This condition is studied for the explicit configurations
in section \ref{sec:For-L+htau},
and the zero-modes are analyzed in detail in appendix 
\ref{sec:anachi}.

In the following, 
we  consider stronger conditions 
which assure the validity of  the commutative limit.
They give a sufficient condition for 
the above condition.  

The first condition for the configuration
$A_i = L_i + \rho a_i$ 
is that the fluctuation $\rho a_i$
should not become as large as the classical background $L_i $.
Otherwise, $\rho a_i$ would change the structure of the space
and violate the assumption that we are considering a gauge theory 
on the fuzzy 2-sphere.  This condition is written as
\beq
|| \ A_{i}^{U} -L_{i} \otimes \mbox{\boldmath  $1$}_{2} \ ||< \epsilon \ , 
~~\epsilon \sim n^0 \ ,
\label{admconupper}
\eeq
for a suitably chosen gauge
$A_{i}^{U}=UA_{i}U^{\dagger}$  
with a unitary matrix  $U$.
Here $|| O ||$  is defined as the maximum value in the absolute values 
of all the eigenvalues of the operator $O$.
Then (\ref{admconupper}) means that all of the eigenvalues are 
bounded by $\epsilon$, which is of the order $n^0$. 

Alternatively to the condition (\ref{admconupper}), we can impose
\beq
|| \ [A_{i} \ , \ A_{j} ]-\epsilon_{ijk} A_{k} \ || < \epsilon' \ ,
~~\epsilon' \sim n^0 \ .
\label{admconupper1}
\eeq
In the commutative limit,
this becomes the condition
that the field strength $F_{ij}$ and the covariant derivative
of the scalar field $D_i \phi$ are bounded by $\epsilon'$.
In this sense, it is similar to 
the admissibility condition in the lattice gauge theory
\cite{Luscher:1981zq,Hernandez:1998et,Luscher:1998kn}\footnote{
A similar admissibility condition on the noncommutative 
torus was studied in \cite{Nagao:2005st}.}. 
The condition will be sufficient if we consider 
fluctuations around the classical background with two-blocks,
but it allows a configuration consisting of more than 2 spheres, 
e.g. a configuration made of 3 irreducible representations of 
$SU(2)$ algebra,
such as $A_i \sim L_i \otimes {\bf 1}_3$.
In order to avoid these configurations, 
we further impose the following condition:
\beq
\left| \ \mbox{Tr} (A_{i}^{2}) -2n \frac{n^{2}-1}{4} \ \right| < \epsilon'' \ , 
~~\epsilon'' \sim o(n^{3}) \ .
\label{admconupper2}
\eeq
Configurations with other-than-two blocks  give 
values of order $n^3$ and hence they are prohibited. 
The monopole configurations (\ref{LnpmLnmm}) 
give values of order $m^2n$.
Thus we have to take $\epsilon''$ smaller than order $n^3$,
and larger than or equal to order $n$.
The condition (\ref{admconupper}) corresponds to
$\epsilon'' \sim n^2$.

The second condition is  that $U(2)$ gauge symmetry
is spontaneously broken to $U(1) \times U(1)$.
Namely, the scalar field must have non-vanishing values 
on arbitrary points on the sphere:
$
\rho^2 \sum_{a=1}^{3}(\phi^a(x))^2  \neq 0
$
for all $x$.
Otherwise we could not define the topological charge (\ref{TCforTP}).
This condition can be satisfied if we impose
\beq
|| \ {\rm tr}_\tau \left(\bigl[(A^U_i)^2-(L_i)^2\bigr] - 
\frac{1}{2}{\rm tr}_\tau \bigl[(A^U_i)^2-(L_i)^2\bigr]\right)^2 \ ||'
\ > \ \eta \ ,
~~\eta \sim n^{2} \ ,
\label{admconlower}
\eeq
in the gauge $A^U_i$ same as in (\ref{admconupper}).
Here $|| O ||'$ means the minimum value in all the eigenvalues
of the operator $O$. 
${\rm tr}_\tau$ stands for a trace over the gauge group space,
leaving matrix components representing sphere coordinates untouched.
Here we used (\ref{A2L2phi}) and (\ref{eqn:(A2L2)2})
to obtain the condition that 
the scalar field in the $SU(2)$ part has non-vanishing values.

These two conditions give 
the lower bound  (\ref{admconlower})
as well as the upper bound (\ref{admconupper}),
or (\ref{admconupper1}) and (\ref{admconupper2}),
on the fluctuations.
While here we considered the conditions that
classical configurations have the appropriate commutative limit,
in order to define quantum theory,
we will need to specify numerical values of the bounds
$\epsilon$ and $\eta$ more precisely.

\subsection{Properties of the topological charge}
\label{sec:topprptopcha}

In this subsection we  consider topological properties 
of the charge (\ref{TCforTP}).
After reviewing some properties in the commutative theory,
we will show that these properties hold in the noncommutative 
theory as well.
In particular, the topological charge is shown to be rewritten as
the winding number of the scalar field in
the noncommutative theory as well as the commutative theory.

The topological charge in the commutative theory is
defined as
the rhs of (\ref{TCforTP}):
\beq
Q_{\rm com} 
=\frac{\rho^2}{8\pi}\int_{S^2} d\Omega \, \epsilon_{ijk} \, n_i
\, {\cal F}_{jk}
\eeq
with
\beq
{\cal F}_{jk}
=\phi'^a F_{jk}^a 
- \epsilon_{abc} \phi'^a (D_j \phi')^b (D_k \phi')^c \ .
\label{comfluxthooft}
\eeq
The flux ${\cal F}_{jk}$ is gauge invariant.
In the $\phi' =(0,0,1)$ gauge, it becomes the flux
in the unbroken $U(1)$ component,
$\partial_j a^3_k - \partial_k a^3_j$.
The charge $Q_{\rm com}$ is also topologically invariant in the sense that
it is invariant under any variations of the gauge fields 
and the scalar field.
One can indeed show that ${\cal F}_{jk}$ is rewritten 
as \cite{Arafune:1974uy}
\beq 
 {\cal F}_{jk}=
 -\epsilon_{abc} \phi'^a (\partial_j \phi'^b) (\partial_k \phi'^c)
 + \partial_j (\phi'^a a^a_k) -  \partial_k (\phi'^a a^a_j) \ .
 \label{comfluxwinding}
\eeq
Then, $Q_{\rm com}$ is equivalent to the winding number 
of the scalar field $\phi'$,
which is known as the Kronecker index,
unless the field configurations have singularities.

In the following, we will show that these properties hold 
in the topological charge of the noncommutative theory, 
the lhs of (\ref{TCforTP}).
In fact, both
${\cal T}r (\Gamma')$  and ${\cal T}r (\hat\Gamma')$ 
are gauge invariant and topologically invariant.
Note that the trace of the sign operator is invariant
under any variations whenever it is changed continuously.
We will study these two quantities,
${\cal T}r (\Gamma')$  and ${\cal T}r (\hat\Gamma')$,
in detail below.

For any configuration $A_{i}=L_{i}+\rho a_{i}$, 
we introduce an interpolating configuration between 
$L_i$ and $A_i$:
\beq
A^h_{i}=L_{i} +h\rho a_{i} \ ,
\label{confAh}
\eeq
where $h$ is a real parameter of ${\cal O}(1)$.
The electric charge operator (\ref{defTgen}) for 
this configuration becomes 
\beq
T'=\frac{h}{|h|}\frac{\{L_{i},\rho a_{i} \}
+h(\rho a_{i})^2}
{\sqrt{\left[ \{L_{i},\rho a_{i} \}+h(\rho a_{i})^2 \right]^{2}  }} \ .
\label{eqn:Tgen}
\eeq
If we restrict our configuration $a_i$ to satisfy
the admissibility conditions, (\ref{admconupper}) and (\ref{admconlower}), 
the eigenvalues of $\left[ \{L_{i},\rho a_{i} \}+h(\rho a_{i})^2 \right]^{2}  $ are
of the order of $n^{2}$ 
while those of $a^{2}_{i}$ are of the order of $1$.  
Thus the denominator of (\ref{eqn:Tgen}) does not have zero-modes
for any $h \sim {\cal O}(1)$.
Then, (\ref{eqn:Tgen}) is a continuous function of $h$,
except for the prefactor $h/|h|$.

$\CTr(\Gamma')$ for this configuration becomes
\beq
\CTr(\Gamma')
=\mbox{Tr}_{R,\sigma}\,(\Gamma)~\mbox{Tr}_{L,\tau}\,(T')
=-2 \frac{h}{|h|}\mbox{Tr}_{L,\tau}\left( \frac{\{L_{i},\rho a_{i} \}
+h(\rho a_{i})^{2}}
{\sqrt{\left[ \{L_{i},\rho a_{i} \}+h(\rho a_{i})^{2} \right]^{2}  }} \right) \ ,
\label{TrGamPgenconh}
\eeq
where $\mbox{Tr}_{R,\sigma}$ denotes a trace of 
matrices which act matrices from the right, and 
over the spinor space.
$\mbox{Tr}_{L,\tau}$ is a trace of matrices which act from the left,
and over the gauge group space.
Since the trace part in (\ref{TrGamPgenconh})
 is topologically invariant,
it takes a constant value for any $h \sim {\cal O}(1)$,
if $a_i$ satisfies the admissibility conditions.
Hence, it can be replaced by the one with $h=0$ as
\beq
\CTr(\Gamma')
=-2 \frac{h}{|h|}\mbox{Tr}_{L,\tau}\,
\left( \frac{\{L_{i},\rho a_{i} \}}{\sqrt{\{L_{i},\rho a_{i} \}^{2}  }} \right) \ .
\label{TrGamma'}
\eeq
Similarly,  
\beq
\CTr(\hat\Gamma')
=\frac{h}{|h|}{\cal T}r \frac{\left\{ \displaystyle{\frac{\{L_{i},\rho a_{i} \}
+h(\rho a_{i})^{2}}{\sqrt{\left[ \{L_{i},\rho a_{i} \}+h(\rho a_{i})^{2} \right]^{2}  }}} \ , 
\ \hat\Gamma  \right\}}
{\sqrt{4+ \left[ \displaystyle{ \frac{\{L_{i},\rho a_{i} \}
+h(\rho a_{i})^{2}}{\sqrt{\left[ \{L_{i},\rho a_{i} \}+h(\rho a_{i})^{2} \right]^{2} } }}  
\ , \ \hat\Gamma \right]^{2}}} \ .
\eeq
is equal to  the one with $h=0$ as
\beq
\CTr(\hat\Gamma')
=\frac{h}{|h|}{\cal T}r \frac{\left\{ \displaystyle{ \frac{\{L_{i},\rho a_{i} \}}
{\sqrt{ \{L_{i},\rho a_{i} \}^{2}  }}} \ ,
\ \frac{2}{n}\left(\sigma \cdot L +\frac{1}{2} \right) \right\}}{\sqrt{4+
  \left[ \displaystyle{\frac{\{L_{i},\rho a_{i} \}}{\sqrt{ \{L_{i}, \rho a_{i} \}^{2}  }}} \ , 
\  \frac{2}{n}\left(\sigma \cdot L +\frac{1}{2} \right) \right]^{2}}} \ .
\label{TrHatGamma'}
\eeq

We now take the commutative limits of (\ref{TrGamma'}) and (\ref{TrHatGamma'}),
and consider their corresponding quantities in the commutative theory.
The commutative limit of (\ref{TrGamma'}) becomes
\beq
\CTr(\Gamma') \to -2\frac{h}{|h|}\, n \int \frac{d\Omega}{4\pi}\,
{\rm tr}_\tau \,\Bigl( 2\phi' + {\cal O}(1/n)\Bigr) \ ,
\label{comlimofTrGamPr}
\eeq 
where ${\rm tr}_\tau$ is a trace over the gauge group space.
The first term vanishes after taking the trace,
if $\phi'$ does not have the $U(1)$ component.
The second term, which is a $1/n$ correction to the scalar field $\phi'$,
gives a finite value,
since  taking ${\rm Tr}_L$ gave a factor $n$ in (\ref{comlimofTrGamPr}).
Therefore, the meaning of (\ref{comlimofTrGamPr}) 
in the commutative theory is obscure.
Although $\CTr(\Gamma')$ is a gauge invariant and topologically
 invariant quantity
in the noncommutative theory,
its commutative counterpart is absent.
This quantity is related to the index of the would-be species-doubler,
as can be shown by the GW algebra \cite{AIM,Fujikawa:1999ku}.

By expanding the denominator of (\ref{TrHatGamma'}), we obtain
\beqa
\CTr(\hat\Gamma')
&=&2 \frac{h}{|h|}\mbox{Tr}_{L,\tau}
\left( \frac{\{L_{i},\rho a_{i} \}}{\sqrt{\{L_{i},\rho a_{i} \}^{2}  }} \right) 
\nonumber \\
&&-\frac{1}{8}\frac{h}{|h|} \left( \frac{2}{n}  \right)^{2} {\cal T}r
\left( \frac{\{L_{i},\rho a_{i} \}}{\sqrt{ \{L_{i},\rho a_{i} \}^{2}  }} (\sigma \cdot L)
 \left[ \frac{\{L_{i},\rho a_{i} \}}{\sqrt{ \{L_{i},\rho a_{i} \}^{2}  }} \ , \ \sigma \cdot L  \right]^{2} \right) 
 +{\cal O}(1/n) \ .
\nonumber \\
\label{TrHatGamma'exp}
\eeqa
The first term is exactly equal to the minus of (\ref{TrGamma'}).
Thus they are canceled in the topological charge,
the lhs of (\ref{TCforTP}).
The second term becomes the winding number of the scalar field $\phi'$
in the commutative limit.
Therefore, in the commutative limit, the topological charge
for the configurations $A_i=L_i + \rho a_i$
becomes the winding number of the scalar field
\beq
\frac{1}{4} {\cal T}r(\Gamma' + \hat\Gamma') 
\to
-\frac{\rho^2}{8\pi} \int_{S^2} d\Omega \
n_i \epsilon_{ijk}
 \epsilon_{abc} \phi'^a (\partial_j \phi'^b) (\partial_k \phi'^c) \ .
\label{topchargecomlimwinding} 
\eeq 
Here we took $h=1$ in order to return the configuration (\ref{confAh})
to the original one $A_i=L_i + \rho a_i$.

In subsection \ref{3.1},
we showed that the commutative limit of the topological charge
becomes that of the commutative theory in (\ref{TCforTP}).
In this subsection, we  have shown that it 
can also be rewritten as  the winding number of 
the scalar field $\phi'$ in (\ref{topchargecomlimwinding}),
by using the topological arguments.
This is consistent with the commutative theory,
shown in (\ref{comfluxthooft}) and (\ref{comfluxwinding}).

\section{Explicit example of configurations}
\label{sec:For-L+htau}
\setcounter{equation}{0}
In this section,
we will consider the following configurations:
\beq
A_{i}=L_{i}+h\frac{\tau_{i}}{2}
\label{confLhtau}
\eeq
for an arbitrary real value $h$.
The $h=1$ case corresponds to the $m=1$ 
TP monopole configuration (\ref{AiLtau}).
We will calculate the topological charge for these configurations.
We also
show that the GW Dirac 
operator can be written as a simple form. 
The results agree with the corresponding commutative cases
if the configurations satisfy the admissibility conditions.
We further extrapolate the configurations to non-admissible regions.

\subsection{Commutative theory}
Before we show calculations in the noncommutative theory,
we study the case in the commutative theory.

From (\ref{defAi}), we see that the gauge field
for (\ref{confLhtau}) is given by
\beq
a_i=\frac{1}{\rho}h\mbox{\boldmath  $1$}_{n} \otimes \frac{\tau_i}{2} \ .
\label{monopoleconfig+h}
\eeq
By taking the commutative limit, and
decomposing it into the tangential and the normal components
on the 2-sphere as in (\ref{decomposeto}), we obtain
\beqa
a_i^{\prime a} &=& h \frac{1}{\rho}\epsilon_{ija} n_j \ , 
\label{TPmonConf-g+h} \\
\phi^a &=& h \frac{1}{\rho} n_a \ .
\label{TPmonConf-h+h}
\eeqa
This is the TP monopole configuration,
(\ref{TPmonConf-g}) and (\ref{TPmonConf-h}),
multiplied by $h$.

As we mentioned in (\ref{comfluxwinding}), 
the topological charge in the commutative theory can be  written as
\beq
Q_{\rm com} = 
-\frac{\rho^2}{8\pi}\int_{S^2} d\Omega \
n_i \epsilon_{ijk}
 \epsilon_{abc} \phi'^a (\partial_j \phi'^b) (\partial_k \phi'^c) \ ,
\label{topchargecomlimwinding+h} 
\eeq 
which is the winding number of the normalized scalar field $\phi'$.
Substituting (\ref{TPmonConf-h+h}), it becomes
\beq
Q_{\rm com} = -\frac{h}{|h|} \ .
\label{Qcomresulth}
\eeq

Furthermore, we can show that the GW Dirac operator 
itself is written simply.
The commutative limit of $D'_{GW}$ is given in (\ref{Dcom}).
As we will show later,
the electric charge operator $T'$ for the configuration 
(\ref{confLhtau}) can be written as (\ref{T'-Ltau}).
Its commutative limit becomes
\beq
T' \to \ \frac{h}{|h|}n \cdot \tau  \ ,
\label{T'-Ltau+h}
\eeq 
where $n_{i}=x_{i}/|x|$ is a unit vector 
for the normal component of the sphere.
Then, from (\ref{comlimTprime}), 
the normalized scalar field is given by
\beq
2 \phi' =\frac{h}{|h|} n \cdot \tau  \ .
\eeq
From (\ref{DGWcom}) and (\ref{monopoleconfig+h}), we obtain
\beqa
D'_{\rm com}
&=& \sigma \cdot {\cal L} +1
+h \frac{1}{2}\Bigl( \sigma \cdot \tau 
-(n \cdot \sigma)(n \cdot \tau) \Bigr) \ .
\eeqa

Therefore, the commutative limit of $D'_{\rm GW}$ becomes
\beqa
\frac{1}{2} \{2\phi' \, , \, D'_{\rm com} \} 
&=& \frac{1}{2} \frac{h}{|h|} \left(  \{ n \cdot \tau \, , \, 
\sigma \cdot {\cal L}+1 \}   
 +h \frac{1}{2}\{ n \cdot \tau \, , \,  \sigma \cdot \tau 
-(n \cdot \sigma)(n \cdot \tau)  \} \right)
\nonumber \\
&=&\frac{h}{|h|} (n \cdot \tau) \  D'^{m=1}_{\rm com} \ ,
\label{D'GW-D'com+h}
\eeqa
where 
\beq
D'^{m=1}_{\rm com}
=\sigma \cdot {\cal L}+1
+\frac{1}{2}\Bigl(\sigma \cdot \tau -
(n \cdot \sigma)(n \cdot \tau)\Bigr) 
\eeq
is the Dirac operator $D'_{\rm com}$ defined by (\ref{DGWcom}),
for the TP monopole configuration
(\ref{monopoleconfig}).
Here we used the relation 
$\{ n \cdot \tau \, , \,  \sigma \cdot \tau 
-(n \cdot \sigma)(n \cdot \tau)  \}=0$.
Owing to this relation, the dependence on $h$ disappeared 
except for the prefactor $h/|h|$ in (\ref{D'GW-D'com+h}).   

In the following subsections, we will show that the same results 
are obtained from the noncommutative theory as well. 

\subsection{Calculations for $\hat\Gamma$}
\label{sec:calhatgamma}

We first note that
we can easily obtain $\frac{1}{4}\CTr (\Gamma' + \hat\Gamma') = -h/ |h|$ 
for the configurations (\ref{confLhtau})
of $h \sim {\cal O}(1)$.
Since $\rho a_i= {\bf 1}_n \otimes \tau_i/2$ satisfies the admissibility conditions,
from the arguments of subsection \ref{sec:topprptopcha},
$\frac{1}{4}\CTr (\Gamma' + \hat\Gamma')$ takes a constant value for 
any $h \sim {\cal O}(1)$.
Moreover, since $\frac{1}{4}\CTr (\Gamma' + \hat\Gamma') = -1$ for $h=1$,
we obtain the above result.
This agrees with the result in the commutative case (\ref{Qcomresulth}).
In the following, we will
perform explicit calculations in the noncommutative theory
for the configurations (\ref{confLhtau})
of an arbitrary real value of $h$,
not restricted to  $h \sim {\cal O}(1)$.

We then consider the chirality operator $\hat\Gamma$ (\ref{gammaRgammahat}).
A crucial observation is that the operator
\beq
H=\sigma \cdot L + \frac{h}{2} \sigma \cdot \tau + \frac{1}{2}
\label{H-sigmaL-hsigmatau}
\eeq
commutes with the total spin operator
\beq
J_{i}=L_{i}+ \frac{\sigma_{i}}{2} + \frac{\tau_{i}}{2} \ ,
\label{JLsigmatau}
\eeq
and thus
\beq
[J_{i} \, , \, \hat\Gamma]=0 \ 
\label{J-hatGamma-commute}
\eeq
is satisfied for an arbitrary real value of $h$. 
It then follows that there exists a simultaneous eigenstate for
$J_i$ and $\hat\Gamma$:
\beqa
(J_{i})^{2}|j, m \rangle
&=&j(j+1) |j, m \rangle \ ,  \\
J_{3}|j, m \rangle&=&m|j, m \rangle \ , \\
\hat\Gamma|j, m \rangle&=&\pm|j, m \rangle \ .
\eeqa
The eigenvalue of $\hat\Gamma$ takes the same value
in each  multiplet of $|j, m \rangle$.


We thus obtain
\beqa
&&\langle j,m| \hat\Gamma(h) |j,m \rangle \nonumber \\
&&=\left(
\begin{array}{ccc}
c_{l+1}(h)\mbox{\boldmath  $1$}_{2l+3} & & \\
 & U(h)\left(
\begin{array}{cc}
c^{1}_{l}(h)  & 0   \\
0 & c^{2}_{l}(h)        
\end{array}
\right)U^{\dagger}(h) 
 \otimes \mbox{\boldmath  $1$}_{2l+1}  &       \\        
   &   &  c_{l-1}(h)\mbox{\boldmath  $1$}_{2l-1}
  \end{array}
\right) \ ,
\nonumber \\
\label{gammahatmatrixrep}
\eeqa
where $c_j(h)$ is the eigenvalue of $\hat\Gamma(h)$ 
in each multiplet $|j,m \rangle$.
Here we introduced $l$ as $n=2l+1$.
Since there is a two-folded degeneracy in $j=l$,
the eigenstate is obtained by 
a unitary transformation $U(h)$
from a fixed basis $|j,m \rangle$.
When the operator $H$ of (\ref{H-sigmaL-hsigmatau}) 
does not have zero-modes,
$\hat\Gamma(h)$ is a continuous function of $h$,
and so are $c_j(h)$. 
Moreover, $c_j(h)$ takes a value of either $1$ or $-1$.
Thus $c_j(h)$ takes a constant value irrespective of $h$.

For $h=0$,
$\hat\Gamma$ is diagonalized by the operator $L_i + \frac{\sigma_i}{2}$,
and we can easily obtain 
$(c_{l+1}, c_l^1, c_l^2, c_{l-1})=(1,1,-1,-1)$.
We can perform similar calculations for 
$h=1, \pm \infty$.
Furthermore, we check zero-modes for the operator 
$H$ of (\ref{H-sigmaL-hsigmatau}).
As we show in Appendix \ref{sec:anachi},
the state of $j=l+1$ becomes a zero-mode at $h=-n$.
The state $j=l-1$ becomes a zero-mode at $h=n$.
The states $j=l$ do not become zero-modes for an arbitrary 
value of $h$.
Consequently, $\hat\Gamma(h)$ has a form of (\ref{gammahatmatrixrep})
with
\beq
(c_{l+1}, c_l^1, c_l^2, c_{l-1})
= \left\{\begin{array}{ll}
(-1,1,-1,-1) & {\rm for} \ h < -n \ , \\
(1,1,-1,-1) & {\rm for} \ -n < h < n \ ,\\
(1,1,-1,1) & {\rm for} \ n < h \ .
\end{array}\right.
\label{gammahateigenvalues}
\eeq
We thus obtain
\beq
\CTr (\hat\Gamma) = 
4n ~~~ {\rm for} \ |h| < n \ , ~~~~
\CTr (\hat\Gamma) = 
2n^2 \frac{h}{|h|}~~~ {\rm for} \ |h| > n \ .
\eeq

We now digress from the calculation for 
$\frac{1}{4} \CTr(\Gamma' + \hat\Gamma')$,
and give some comments on
a naive topological charge without introducing 
the projection operator
or the electric charge operator.
It becomes 
\beqa
\frac{1}{2}\CTr(\Gamma+\hat\Gamma) &=& 
0~~~~ {\rm for} \ |h| < n \ , \label{naivetopch0}\\
\frac{1}{2}\CTr(\Gamma+\hat\Gamma) &=& 
n^2 \frac{h}{|h|}-2n~~~~ {\rm for} \ |h| > n \ .
\label{naivetopchnonadm}
\eeqa
From the same arguments of subsection \ref{sec:topprptopcha},
for any admissible configurations,
$\frac{1}{2}\CTr(\Gamma+\hat\Gamma)$
takes the same value as for the case of $a_i =0$,
and thus we have $\frac{1}{2}\CTr(\Gamma+\hat\Gamma)=0$.
The result (\ref{naivetopch0}) agrees with this fact. 
Moreover, (\ref{naivetopch0}) shows that this result
is kept in quite large regions of the configuration space,
even in the non-admissible regions of $h \sim n$.
On the other hand,
(\ref{naivetopchnonadm}) shows that the topological charge 
can take nonzero values
for non-admissible configurations.

In fact, the topological charge $\frac{1}{2}\CTr(\Gamma+\hat\Gamma)$
takes various integer values 
for various matrix configurations,
while only the topologically-trivial sector remains after 
imposing the admissibility conditions.
The same results were obtained in the noncommutative torus \cite{ANS}.
This situation is in striking contrast to the commutative case.
In the ordinary lattice gauge theories,
all of the topological sectors remain
even after imposing the admissibility conditions.
This discrepancy can be explained as follows:
Configurations with nontrivial topologies are described 
in three ways.
The first way is to consider a singular configuration
and put nontriviality on the singularity.
The second one is to consider 
the theory with the twisted boundary conditions
or to introduce the notion of patch.
The third one is to use the spontaneous symmetry-breakdown
of the gauge symmetry.
However,
noncommutative geometry smears out singularities of configurations, 
and prohibits the first way.
Thus only the  trivial topological sector can exist if one specifies 
the  trivial boundary conditions.

This is the case when we consider the naive topological charge
$\frac{1}{2}\CTr(\Gamma+\hat\Gamma)$.
We can obtain a nontrivial topology within the admissible configurations
if we introduce the projection operator as in
section \ref{sec:ITforTP}.
Noncommutative gauge theory with the twisted boundary conditions
is also formulated by the finite size matrix model
in \cite{ANS2}.
Furthermore, as we showed in section \ref{sec:Form-gen},
we can define all of the topological sectors from a single theory,
by describing the topology in terms of the winding number of the scalar field
in the spontaneously symmetry-broken gauge theory.

\subsection{Calculations for $\hat\Gamma'$}

\begin{figure}[ht]
\begin{center}
\includegraphics[width=15cm,height=6cm]{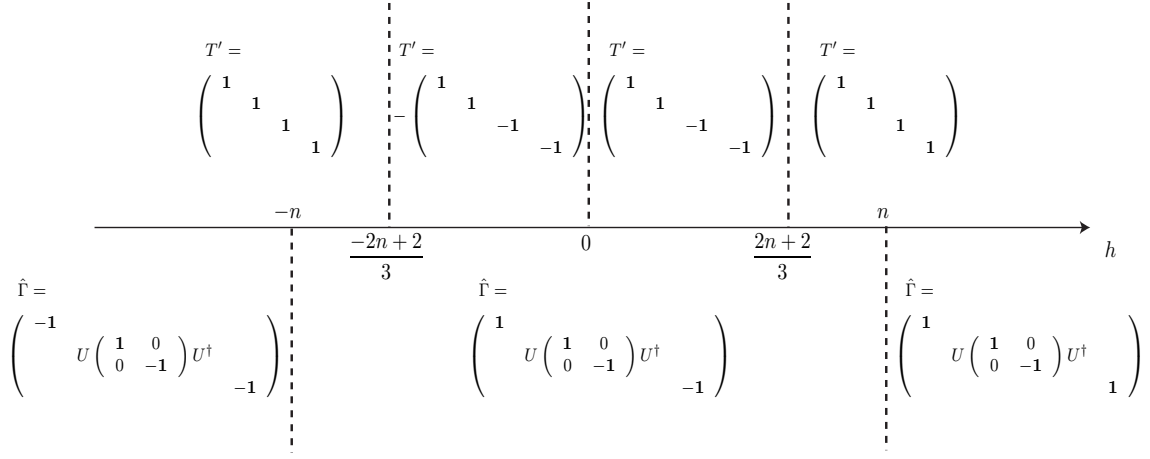}
\caption{The forms of $\hat\Gamma$ and $T'$as functions of $h$,
(\ref{gammahateigenvalues}), 
(\ref{T'-Ltau}) and (\ref{T'ohterregion}).
Here the bases in the $j=l$ sector are taken as the eigenstates for
the operator $L_i+\tau_i/2$.
}
\label{h-TG}
\end{center}
\end{figure}

We now consider the electric charge operator $T'$ of (\ref{defTgen})
for the configurations (\ref{confLhtau}). 
Using
\beq
(A_{i})^{2}-(L_{i})^{2}
=h L \cdot \tau +\frac{3}{4} h^{2} \ ,
\eeq
we obtain 
\beq
T'=\frac{h}{|h|}
\frac{L \cdot \tau +\displaystyle{\frac{3}{4}} h}
{\sqrt{\left( L \cdot \tau +\displaystyle{\frac{3}{4}} h \right)^2}} \ .
\label{T-Lt}
\eeq
For the $(n \pm 1)$ dimensional irreducible representation 
of $L_{i}+\tau_{i}/2$, 
the operator $L \cdot \tau$ takes the following values: 
\beqa
L \cdot \tau
&=&\left( L_{i}+\frac{\tau_{i}}{2} \right)^{2}-(L_{i})^{2}-
\left(\frac{\tau_{i}}{2} \right)^{2} 
\nonumber \\
&=&\left\{
\begin{array}{ll}
\displaystyle{\frac{2n-2}{4}}  & \mbox{for $(n+1)$ dim. rep.}   \\ 
\displaystyle{\frac{-2n-2}{4}} & \mbox{for $(n-1)$ dim. rep.}        
\end{array}
\right.
\eeqa
In the $(n+1)$ dimensional representation, 
we obtain $T'=\frac{h}{|h|}\frac{2n-2+3h}{|2n-2+3h|}$,
and then $T'=\frac{h}{|h|}$ for $h > (-2n+2)/3$.
In the $(n-1)$ dimensional representation, 
we have $T'=\frac{h}{|h|}\frac{-2n-2+3h}{|-2n-2+3h|}$,
and $T'=-\frac{h}{|h|}$ for $h < (2n+2)/3$.
Therefore, $T'$ is written as
\beqa
T'&=&\frac{h}{|h|}\, \frac{2}{n}\left(L \cdot \tau + \frac{1}{2} \right) 
~~~~{\rm for}~\frac{-2n+2}{3} <h < \frac{2n+2}{3} \ ,
\label{T'-Ltau} \\
T'&=& {\bf 1} ~~~~
{\rm for  \ the \ other \ regions \ of} \ h \ .
\label{T'ohterregion}
\eeqa
Moreover, $T'$ commutes with the total spin  operator $J_i$
defined in (\ref{JLsigmatau}),
and then (\ref{T'-Ltau}) is rewritten as a form of (\ref{gammahatmatrixrep})
with $(c_{l+1}, c_l^1, c_l^2, c_{l-1})=(1,1,-1,-1)$
multiplied by $\frac{h}{|h|}$.
We summarize the forms of $T'$ and $\hat\Gamma$ in
Figure \ref{h-TG}.

We next consider the modified chirality operator $\hat\Gamma'$
of (\ref{defgammahatprime}).
For $h > (2n+2)/3$ and for $h < (-2n+2)/3$,
$\hat\Gamma'$ reduces to $\hat\Gamma$, 
whose form was already given in (\ref{gammahateigenvalues}).
For $(-2n+2)/3 <h <(2n+2)/3$,
zero-modes might occur from the anti-commutator
of $T'$ and $\hat\Gamma$ in the $j=l$ sector.
As we show in Appendix \ref{sec:anachi}, 
such zero-modes do not take place.
Consequently, we obtain 
\beq
\langle j,m| \hat\Gamma'(h) |j,m \rangle 
=\frac{h}{|h|} \left(
\begin{array}{cccc}
\mbox{\boldmath  $1$}_{2l+3} & & & \\
 & -\mbox{\boldmath  $1$}_{2l+1}  & &       \\        
 && -\mbox{\boldmath  $1$}_{2l+1}   &       \\        
 & &&  \mbox{\boldmath  $1$}_{2l-1}
  \end{array}
\right) \ 
\label{gammahatprime1-1-11}
\eeq
for $(-2n+2)/3 <h <(2n+2)/3$.
It is independent of $h$ except for the prefactor $h/|h|$.

\begin{figure}[hp]
\begin{center}
\includegraphics[width=14.5cm,height=4.2cm
]{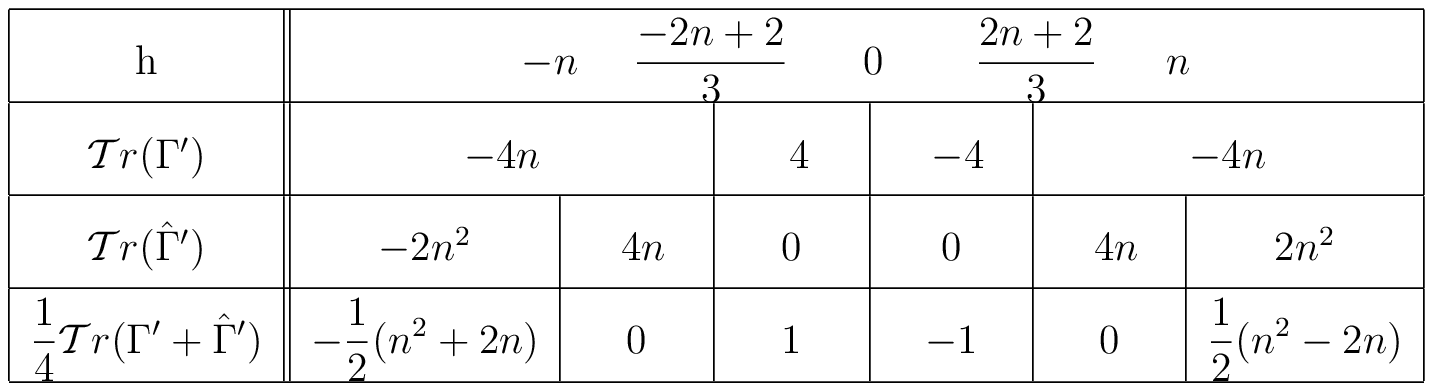}
\caption{Topological charge 
$\frac{1}{4} {\cal T}r [\Gamma' + \hat\Gamma']$
of the configurations (\ref{confLhtau})
for an arbitrary real value $h$.
}
\label{h-TC}
\vspace{1.5cm}
\includegraphics[width=12cm
]{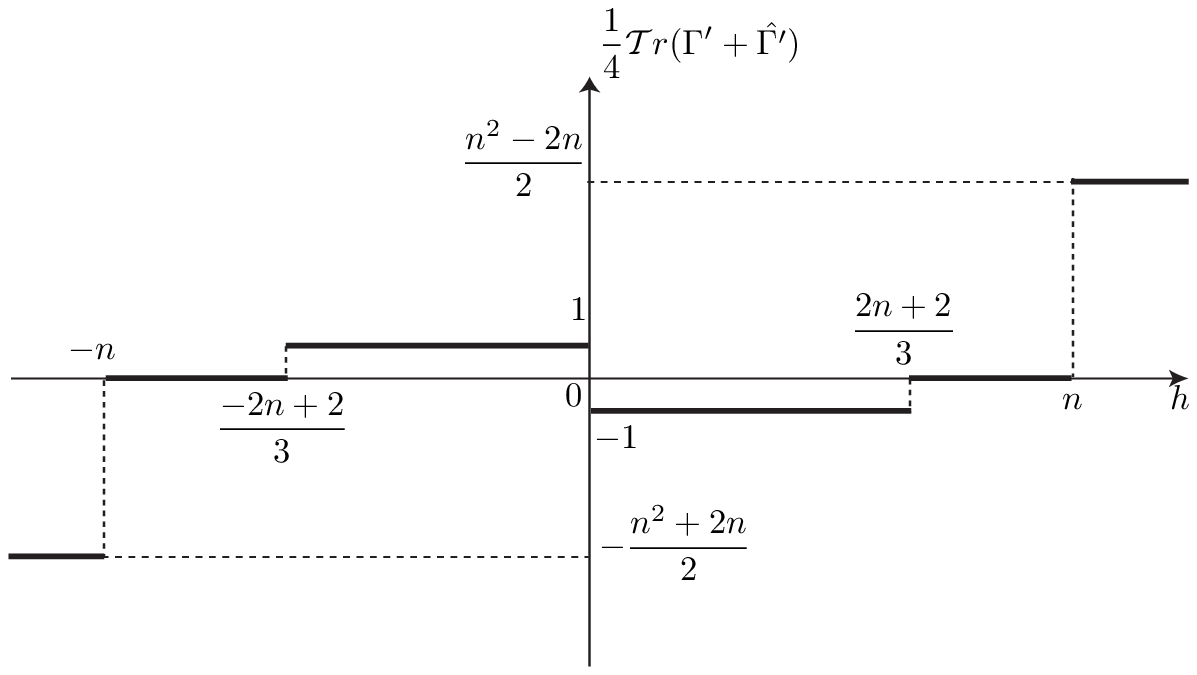}
\caption{Topological charge
$\frac{1}{4} {\cal T}r [\Gamma' + \hat\Gamma']$
as a function of $h$.
The $h=1$ case corresponds to the TP monopole configuration
of the previous formulation.
In the admissible regions $|h| \sim 1$,
the results agree with the commutative case.
We further obtained results for an arbitrary real value of $h$,
extending to non-admissible regions.
}
\label{TC}
\end{center}
\end{figure}

Now that we have evaluated the operators $T'$, $\hat\Gamma'$,
we can easily evaluate $\frac{1}{4}\CTr(\Gamma'+ \hat\Gamma')$.
The results are shown in Figure \ref{h-TC} and in Figure \ref{TC}.
In particular, we obtain
\beq
\frac{1}{4} {\cal T}r [\Gamma' + \hat\Gamma']
=-\frac{h}{|h|} \ ,
\label{Trgammaprimehoverh}
\eeq
for $(-2n+2)/3 <h <(2n+2)/3$.
This agrees with the result in the commutative theory 
(\ref{Qcomresulth}).
As we mentioned at the beginning of
subsection \ref{sec:calhatgamma},
this result can also be obtained from the arguments of 
subsection \ref{sec:topprptopcha} for 
$h \sim {\cal O}(1)$.
Note also that we obtain both positive and negative topological charge,
while we could define only negative charge in the previous formulation,
as we pointed out in (\ref{topcharnegative-m}). 
Furthermore, we have obtained the results for non-admissible regions 
as well.
The result (\ref{Trgammaprimehoverh}) holds even in the
non-admissible regions of $h \sim n$,
but it changes its value if we further extend the 
value of $h$.

Since the chirality operator $\hat\Gamma'$, 
(\ref{gammahatprime1-1-11}),
is independent of $h$ except for the prefactor,
it can be written as 
\beq
\hat\Gamma'(h)
=\frac{h}{|h|} \hat\Gamma' (h=1) 
=T'(h) \hat\Gamma(h=1) \ .
\label{hatGamma'-h}
\eeq
The other chirality operator $\Gamma'$ is written as
\beq
\Gamma'(h)
=T'(h) \Gamma \ .
\label{Gamma'-h}
\eeq
Therefore,
the GW Dirac operator (\ref{defDGWgen}) reduces to 
\beq
D'_{GW}=T'(h) D^{m=1}_{GW} \ ,
\label{D'GWh-T'DGW}
\eeq
for $(-2n+2)/3 <h <(2n+2)/3$.
Here $D^{m=1}_{GW}$ is the GW Dirac operator 
of the previous definition (\ref{defDGW}),
for the TP monopole configuration  
with magnetic charge $m=1$ (\ref{AiLtau})\footnote{
Incidentally, $D^{m=1}_{GW}$ can be written as
$
D^{m=1}_{GW}
=\sigma \cdot {\tilde L} +1 +\frac{1}{2} \sigma \cdot \tau
-\frac{1}{n^{2}-1} L \cdot \tau
\left[ 1+2\sigma \cdot \left( L + \frac{\tau}{2}  \right) \right] \ .
$
.}.
$D^{m=1}_{GW}$ is independent of $h$.
$T'(h)$ is given in (\ref{T'-Ltau}).
This result (\ref{D'GWh-T'DGW}) agrees with
the commutative case (\ref{D'GW-D'com+h}).

Now that we have obtained the simple form for 
the GW Dirac operator (\ref{D'GWh-T'DGW}),
we can easily calculate various quantities, such as the spectrum of
the Dirac operator for the configurations (\ref{confLhtau}),
as was done for the configurations 
(\ref{LnpmLnmm}) in \cite{AIM}.

\section{Conclusions and Discussions}
\label{conclusion}
\setcounter{equation}{0}

In this paper,
we studied the topological structure of spontaneously symmetry-broken
gauge theory on the fuzzy 2-sphere,
by examining
the index theorem which is applicable to general configurations,
not restricted to a special type of configurations.
We showed in detail that
the commutative limit of the topological charge
becomes the appropriate one introduced by 't Hooft.
Since this formulation is valid for general configurations,
 configuration space can be classified
into topological sectors.

We then discussed the conditions to assure the validity 
of this formulation,
which gave both upper and lower bounds to the fluctuations,
though the ordinary admissibility condition in the lattice 
gauge theory gives only the upper bound.
It is  an interesting future problem 
to devise  a mechanism which dynamically
realizes these conditions rather than imposing them by hand.
For example, we can deform the bosonic action to
prevent configurations which are prohibited 
by these conditions \cite{Luscher:1998kn}.
This will open possibilities to perform Monte Carlo simulations
of this formulation.

We also studied some topological properties of the topological charge.
In particular, we showed that the topological charge is rewritten 
as the winding number of the scalar field in the noncommutative theory
as well as in the commutative theory.
We also found the gauge invariant and topologically invariant quantity,
$\CTr (\Gamma')$,
in the noncommutative theory,
whose counterpart in the commutative theory is absent.
This quantity is related to the index of the would-be species-doubler.
Although it is an analogue of a lattice artifact, it plays an important role
in defining the index consistently in theories with finite
degrees of freedom.

We further investigated some explicit configurations.
We calculated the topological charge and obtained the simple form of 
the GW Dirac operator 
for these configurations.
The results agree with the commutative case
if the configurations satisfy the admissibility conditions.
We also showed that we can define both positive and negative
topological charge, while in the previous formulation we could only
define the negative charge.
We further obtained the results for non-admissible regions.
Furthermore, since we obtained the explicit form of the Dirac operator
for these configurations, we can calculate various quantities
as the spectrum of the Dirac operator.
While here we studied a series of configurations which connect
the topological sectors with the topological charge $1$ and $-1$,
it is also interesting to study configurations of other sectors. 
In order to study the configurations with the topological charge 
greater than or equal to $2$, however, it may be necessary to 
 obtain another representation of the 
configurations where space and gauge field are written separately.

We finally give some comments on the implications to
the string theory compactifications.
While in the general formulation studied in this paper,
all of the topological sectors are defined from a single theory
as in the commutative theories,
defining the projective module in the noncommutative theory
gives only a single topological sector,
as in the previous formulation for the TP monopoles
using the projection operator.
This feature of the noncommutative theory,
if we use it in the compactified spaces in the string theories,
may be useful to determine the number of matter generations.
On the other hand, we have observed 
various topological sectors in the non-admissible regions,
from the calculations for the explicit configurations.
They are different from the so-called noncommutative solitons and fluxons
\cite{ncsol},
which are also new topological objects in the noncommutative theory
but appear only in the single topological sector specified by the theory.
In the compactified spaces in the string theories,
the notion of the ordinary space may be spoiled 
and the description of finite size matrices 
may become more appropriate.
In these cases, the novel topologies in the non-admissible regions
may play an important role in determining
matter contents on our spacetime.
We hope that the finite matrix will give  
a new possibility for the string theory compactifications.


\appendix

\section{Commutative limit of the electric charge operator}
\label{sec:U1}

In this appendix, we show that 
the electric charge operator $T'$ becomes 
the normalized scalar field $\phi'$ 
in the commutative limit,
(\ref{comlimTprime}).

Since we consider the $U(2)$ gauge group, 
the gauge field has the $SU(2)$ part and the $U(1)$ part as
\begin{equation}
A_{i}=L_{i}+
\rho\left(a^{a}_{i}\frac{\tau^{a}}{2}+a^{0}_{i}\frac{\bf 1}{2}\right) \ ,
\label{eqn:ALaa}
\end{equation}
where the first term $L_i$ is of ${\cal O}(n)$.
Here we will assume that the $SU(2)$ part $\rho a^{a}_{i}$ is of 
order one
and the $U(1)$ part $\rho a^{0}_{i}$ is of ${\cal O}(1/n)$.

Then, the scalar field 
\begin{equation}
(A_i)^2-(L_i)^2
=
\rho n \left(\phi^{a} \frac{\tau^{a}}{2}
+\phi^{0}\frac{\bf 1}{2}\right)
\label{A2L2phi}
\end{equation}
has the $SU(2)$ part $\rho\phi^{a}$ of order one
and the $U(1)$ part $\rho\phi^{0}$ of ${\cal O}(1/n)$.
The square of (\ref{A2L2phi}) becomes
\begin{equation}
\left[ (A_i)^2-(L_i)^2  \right]^{2}
= 
\frac{1}{4}\rho^2n^2\left[ i \epsilon_{abc} [\phi^{a}\, , \, \phi^{b}]\tau^{c}
+(\phi^{a})^{2}+\{\phi^{a} \, , \, \phi^{0} \}\tau^a+(\phi^{0})^{2} \right] \ .
\label{eqn:(A2L2)2}
\end{equation}
In this form,
the second term $\rho^2(\phi^{a})^{2}$ is of order one
and the other terms are of ${\cal O}(1/n)$.

Therefore, in the commutative limit, 
the electric charge operator $T'$ becomes
\beq
T' = \frac{(A_i)^2-(L_i)^2}{\sqrt{[(A_i)^2-(L_i)^2]^2}}
\, \to \,
\frac{2 \phi^{a}(x) \displaystyle{\frac{\tau^{a}}{2}} }{\sqrt{(\phi^{a}(x))^{2} }}
=2 \, \phi'^{a}(x)\frac{\tau^{a}}{2}
=2 \, \phi' (x) \ ,
\eeq
and (\ref{comlimTprime}) is shown.

The reason why we assumed that the $U(1)$ part is negligibly small 
is for the normalization of $\phi'$.
This assumption has nothing to do with the admissibility conditions,
discussed in subsection \ref{sec:admissibilitycon},
which assure the validity of the formulation.
It is then desirable to define a more elaborate $T'$, which has
the proper commutative limit without any constraints to
the configurations.

\section{Commutative limit of the Dirac operator and the
topological charge}
\label{sec:comlimtopcharge}
\setcounter{equation}{0}

In this appendix, we show the calculations of
taking the commutative limit of the Dirac operator, (\ref{Dcom}), 
and the topological charge, (\ref{TCforTP}).

The denominator of the chirality operator $\hat\Gamma'$
can be written as
\beq
\{T' \, , \, \hat\Gamma\}^2=4+[T' \, , \, \hat\Gamma]^2 \ ,
\label{anticomcom}
\eeq
since $(T')^2=1$ and $(\hat\Gamma)^2=1$.
The second term in (\ref{anticomcom})
is of the order of $1/n^2$.
We thus obtain
\beq
\hat\Gamma' =\frac{1}{2}\{T' \, , \, \hat\Gamma\} 
-\frac{1}{16}\{T' \, , \, \hat\Gamma\}[T'\, , \, \hat\Gamma]^2 
+{\cal O}\left(\frac{1}{n^3} \right) \ .
\label{gammahatprimeexpand}
\eeq

For taking the commutative limit of the 
Dirac operator $D'_{\rm GW}$,
it is enough to take the first term in 
(\ref{gammahatprimeexpand}) into account.
We can easily see 
\beqa
D'_{\rm GW} &=& a^{-1} \frac{1}{2}\{T',(\hat\Gamma-\Gamma)\} 
+{\cal O}(1/n) \\
&\to& \frac{1}{2}\{2\phi' \, , \, D'_{\rm com}\} \ .
\eeqa
In particular, in the $\phi'^a(x)=(0,0,1)$ gauge, it becomes
the Dirac operator with the coupling to 
the unbroken $U(1) \times U(1)$ gauge fields.

For taking the commutative limit of the topological charge,
however,
we have to take into account the second term in 
(\ref{gammahatprimeexpand}) as well.
We then have
\beq
\frac{1}{4}{\cal T}r[\Gamma' +\hat{\Gamma}']
=\frac{1}{4}{\cal T}r \left[ T'(\Gamma+\hat{\Gamma})
-\frac{1}{8} T' \hat{\Gamma}[T' \, , \, \hat\Gamma]^2 
+{\cal O}\left( \frac{1}{n^3} \right) \right] \ .
\eeq
Note that the first and the second terms are of the order of $1/n^2$,
and give finite values after taking the trace,
since taking trace gives a factor $n^2$.
The first term becomes in the commutative limit
\beq
\frac{1}{4}{\cal T}r[ T'(\Gamma+\hat{\Gamma})]
\, \to \, \frac{\rho^2}{8\pi}\int_{S^2} d\Omega \ \epsilon_{ijk}
n_i \phi'^a F_{jk}^a 
\eeq
as in \cite{AIN2,AIN3}.
$F_{jk}=F_{jk}^a \tau^a/2$ 
is the field strength defined as
$F_{jk}= \partial_j a_k'-\partial_k a_j'-i[a_j',a_k']$,
where $a'_{i}=\epsilon_{ijk} x_{j} a_{k}/ \rho $ is the tangental components 
of the gauge field on the sphere.
The second term becomes 
\beq
-\frac{1}{4}{\cal T}r\left[\frac{1}{2} T' 
\hat{\Gamma} \left[\hat\Gamma \, , \, \frac{1}{2} T' \right]^2 \right]
\, \to \,
-\frac{1}{4} \frac{n^2}{4 \pi} \int_{S^2} d\Omega  \ \tr_{\sigma,\tau}
\Bigl[\phi' (n\cdot \sigma)
\Bigl(-ia \rho  \epsilon_{ijk} \sigma_i n_j( D_k \phi')\Bigr)^2
\Bigr] \ ,
\label{TC2}
\eeq
where $\tr_{\sigma,\tau}$ is a trace over
the spinor space and over the gauge group space.
$D_{j}$ 
is the covariant derivative operator defined as
$D_{j} =\partial_{j} -i[a'_{j} , ~~]$.
Here we used
\beq
\left[\hat\Gamma \, , \, \frac{1}{2} T' \right]
\to -ia \rho \epsilon_{ijk} \sigma_{i} n_{j} (D_{k} \phi') \ . 
\eeq
Taking the trace $\tr_{\sigma,\tau}$, (\ref{TC2}) becomes
\beq
-\frac{\rho^2}{8\pi}\int_{S^2} d\Omega \ \epsilon_{ijk}
n_i 
 \epsilon_{abc} \phi'^a (D_j \phi')^b (D_k \phi')^c \ .
\eeq
Therefore, the commutative limit of the topological charge becomes
\beq
\frac{1}{4}{\cal T}r[\Gamma' +\hat{\Gamma}']
\, \to \,
\frac{\rho^2}{8\pi}\int_{S^2} d\Omega \ \epsilon_{ijk}
n_i \Bigl( \phi'^a F_{jk}^a 
- \epsilon_{abc} \phi'^a (D_j \phi')^b (D_k \phi')^c \Bigr) \ ,
\eeq
which is precisely the topological charge introduced by 't Hooft 
\cite{'tHooft:1974qc}.

\section{Analyses on zero-modes in the chirality operators}
\label{sec:anachi}
\setcounter{equation}{0}

In this appendix,
we investigate zero-modes in the chirality operators 
$\hat\Gamma$ and  $\hat\Gamma'$
to obtain the results (\ref{gammahateigenvalues}) 
and (\ref{gammahatprime1-1-11}).

\subsection{Bases and unitary transformations}

As we mentioned below (\ref{gammahatmatrixrep}),
there exists a two-folded degeneracy in the $j=l$ sector.
We thus have an ambiguity to choose two multiplets,
which we will call
$|+ \rangle $ and $|-\rangle$.
We here introduce three types of the bases $| \pm \rangle_{i}$
with $i=1,2,3$:
$| \pm \rangle_{1}$ are diagonalized by a spin operator 
$L_{i}+\frac{\tau_{i}}{2}$
and have spin $l \pm \frac{1}{2}$ respectively.
Similarly, $| \pm \rangle_{2}$ are diagonalized by 
$L_{i}+\frac{\sigma_{i}}{2}$ with spin $l \pm \frac{1}{2}$.
$| \pm \rangle_{3}$ are diagonalized by
$\frac{\sigma_{i}}{2}+\frac{\tau_{i}}{2}$
with spin $1/2 \pm 1/2$. 
Therefore, these states are eigenstates for
the following operators:
\beqa
L \cdot \tau
&=&\left\{
\begin{array}{ll}
l=
\displaystyle{\frac{n-1}{2}}   & \mbox{for $|+ \rangle_{1} $} \ ,   \\ 
-(l+1)=
\displaystyle{-\frac{n+1}{2}}  & \mbox{for $|- \rangle_{1} $} \ ,        
\end{array}
\right.  
\label{LdotTAU} \\
L \cdot \sigma
&=&\left\{
\begin{array}{ll}
l=
\displaystyle{\frac{n-1}{2}}   & \mbox{for $|+ \rangle_{2} $} \ ,   \\ 
-(l+1)=
\displaystyle{-\frac{n+1}{2}}  & \mbox{for $|- \rangle_{2} $} \ ,        
\end{array}
\right. 
\label{LdotSIGMA} \\
\sigma \cdot \tau
&=&\left\{
\begin{array}{ll}
1  & \mbox{for $|+ \rangle_{3} $} \ ,   \\ 
-3  & \mbox{for $|- \rangle_{3} $} \ .        
\end{array}
\right.
\label{SIGMAdotTAU}  
\eeqa

Different types of bases are related to one another 
by unitary transformation 
\begin{equation} 
| a \rangle_{i} = \sum_{b= \pm} U^{(ij)}_{ab}| b \rangle_{j} \ .
\end{equation}
The unitary matrices $U^{(ij)}$ have the following forms:
\begin{eqnarray}
U^{(12)}
&=&\left(
\begin{array}{cc}
\displaystyle{\frac{1}{2l+1}}  & -\displaystyle{\frac{2\sqrt{l(l+1)}}{2l+1}}   \\
\displaystyle{\frac{2\sqrt{l(l+1)}}{2l+1}} & \displaystyle{\frac{1}{2l+1}}       
\end{array}
\right) \ ,  \\
U^{(23)}
&=&\left(
\begin{array}{cc}
\displaystyle{\sqrt{\frac{l}{2l+1}}}  & -\displaystyle{\sqrt{\frac{l+1}{2l+1}}}    \\
\displaystyle{\sqrt{\frac{l+1}{2l+1}}}  & \displaystyle{\sqrt{\frac{l}{2l+1}}}        
\end{array}
\right) \ ,  \label{U23}  \\
U^{(13)}
&=&\left(
\begin{array}{cc}
-\displaystyle{\sqrt{\frac{l}{2l+1}}}  & -\displaystyle{\sqrt{\frac{l+1}{2l+1}}}    \\
\displaystyle{\sqrt{\frac{l+1}{2l+1}}}  & -\displaystyle{\sqrt{\frac{l}{2l+1}}}        
\end{array}
\right) \ .
\end{eqnarray}

This can be checked by comparing the highest-weight state,
namely the state with $j=l, j_z=l$,
in each multiplet:
\begin{eqnarray*}
&&\left\{ \begin{array}{l}
\displaystyle{ |+\rangle_{1} =
\sqrt{\frac{l}{(l+1)(2l+1)}} ~| l-1 \uparrow \uparrow \rangle
+\sqrt{\frac{1}{2(l+1)(2l+1)}} ~| l \uparrow \downarrow \rangle
-\sqrt{\frac{2l+1}{2(l+1)}} ~| l  \downarrow \uparrow \rangle  \ , }
\\
\displaystyle{
|-\rangle_{1} =
-\sqrt{\frac{1}{2l+1}} ~| l-1 \uparrow \uparrow \rangle
+\sqrt{\frac{2l}{2l+1}} ~| l \uparrow \downarrow \rangle  \ , }
\end{array}
\right. \\
&&\left\{ \begin{array}{l}
\displaystyle{
|+\rangle_{2} =
-\sqrt{\frac{l}{(l+1)(2l+1)}} ~| l-1 \uparrow \uparrow \rangle
+\sqrt{\frac{2l+1}{2(l+1)}} ~| l \uparrow \downarrow \rangle
-\sqrt{\frac{1}{2(l+1)(2l+1)}} ~| l  \downarrow \uparrow \rangle  \ , }
\\
\displaystyle{
|-\rangle_{2} =
-\sqrt{\frac{1}{2l+1}} ~| l-1 \uparrow \uparrow \rangle
+\sqrt{\frac{2l}{2l+1}} ~| l  \downarrow \uparrow \rangle  \ , }
\end{array}
\right. \\
&&\left\{ \begin{array}{l}
\displaystyle{
|+\rangle_{3} =
-\sqrt{\frac{1}{l+1}} ~| l-1 \uparrow \uparrow \rangle
+\sqrt{\frac{l}{2(l+1)}} ~| l \uparrow \downarrow \rangle
+\sqrt{\frac{l}{2(l+1)}} ~| l  \downarrow \uparrow \rangle \ , }
\\
\displaystyle{
|-\rangle_{3} =
-\frac{1}{\sqrt{2}} ~| l \uparrow \downarrow \rangle
+\frac{1}{\sqrt{2}} ~| l  \downarrow \uparrow \rangle \ . }
\end{array}
\right. \\
\end{eqnarray*}

\subsection{Calculations for zero-modes in $\hat\Gamma$}

We now study zero-modes for the operator 
$H$ of (\ref{H-sigmaL-hsigmatau}).
The zero-mode equation $H |\psi \rangle =0$ is 
written as
\beq
\left( \sigma \cdot L +\frac{1}{2}  \right) | \psi \rangle
=-h\frac{1}{2} \sigma \cdot \tau |\psi \rangle \ .
\label{zeromode}
\eeq
The state of $j=l+1$ is a simultaneous eigenstate
for the operators in both sides of (\ref{zeromode}).
The lhs gives $\frac{n}{2}$,
while the rhs gives $-\frac{h}{2}$.
Therefore $j=l+1$ state becomes a zero-mode at $h=-n$.
Similarly,  the state $j=l-1$ becomes a zero-mode at $h=n$.

For the states $j=l$, we consider a linear combination
$|\psi \rangle =c_{+} | +\rangle_{2}+c_{-} | -\rangle_{2}$. 
We here took the basis $| \pm \rangle_{2}$.
From (\ref{LdotSIGMA}) and (\ref{SIGMAdotTAU}),
(\ref{zeromode}) is written as
\beq
\left[
\frac{n}{2}
\left(
\begin{array}{cc}
1  & 0   \\
0 & -1        
\end{array}
\right)
+\frac{h}{2} \ (U^{23})^{*} 
\left(
\begin{array}{cc}
1  & 0   \\
0 & -3        
\end{array}
\right)
(U^{23})^{T} 
  \right]
\left(
\begin{array}{c}
c_{+}     \\
c_{-}         
\end{array}
\right)=0 \ .
\label{zeromodesjeqlsec}
\eeq 
By taking $U^{(23)}$ as
\beq
U^{(23)}=
\left(
\begin{array}{cc}
\cos \theta  & -\sin \theta   \\
\sin \theta  & \cos \theta        
\end{array}
\right) \ ,
\label{U23-cs}
\eeq
(\ref{zeromodesjeqlsec}) becomes
\beq
\frac{1}{2}
\left(
\begin{array}{cc}
n/h +1 -4\sin^2\theta  & 4\sin\theta \cos\theta  \\
4\sin\theta \cos\theta & -n/h +1 -4\cos^2\theta       
\end{array}
\right)
\left(
\begin{array}{c}
c_{+}     \\
c_{-}         
\end{array}
\right)=0 \ .
\eeq
This equation has a nontrivial solution if and only if 
\beq
\left( \frac{n}{h} \right)^{2} +4\cos (2\theta) \frac{n}{h} +3=0 
\eeq
is satisfied.
This is satisfied by some real value $h$
if $\cos^{2} (2\theta) \ge 3/4$.
On the other hand, 
by comparing (\ref{U23-cs}) with (\ref{U23}), we have
\beq
\cos \theta =\sqrt{\frac{l}{2l+1}},
~ \sin\theta = \sqrt{\frac{l+1}{2l+1}} \ ,
\eeq
and thus we obtain
\beq
\cos^{2} (2\theta)=\frac{1}{(2l+1)^2} =\frac{1}{n^{2}} 
< \frac{3}{4} 
\eeq
for $n \ge 2$.
Consequently, the states $j=l$ do not have zero-modes
for an arbitrary real value $h$ 
if  $n \ge 2$.

Therefore, we obtain our result
(\ref{gammahateigenvalues}).

\subsection{Calculations for zero-modes in $\hat\Gamma'$}

We next consider zero-modes in $\hat\Gamma'$. 
From Figure \ref{h-TG},
we see that
for $(-2n+2)/3 <h <(2n+2)/3$,
$T'$ and $\hat\Gamma(h)$  are written as
\beqa
\langle j,m| T' |j,m \rangle 
&=&\frac{h}{|h|} \left(
\begin{array}{cccc}
\mbox{\boldmath  $1$}_{2l+3} & & & \\
 & \mbox{\boldmath  $1$}_{2l+1}  & &       \\        
 && -\mbox{\boldmath  $1$}_{2l+1}   &       \\        
 & &&  -\mbox{\boldmath  $1$}_{2l-1}
  \end{array}
\right) \ , \label{T}  \\    
\langle j,m| \hat\Gamma(h) |j,m \rangle 
&=& \left(
\begin{array}{ccc}
\mbox{\boldmath  $1$}_{2l+3} & &  \\
 & U(h)\left(
\begin{array}{cc}
1  & 0   \\
0 & -1        
\end{array}
\right)U^{\dagger}(h)
\otimes \mbox{\boldmath  $1$}_{2l+1}   &       \\        
 & &  -\mbox{\boldmath  $1$}_{2l-1}
  \end{array}
\right) \ .  \nonumber \\
\eeqa
We here took the bases $| \pm \rangle_1$.
The unitary matrix $U(h)$ relates $| \pm \rangle_1$
to  $| \pm \rangle_H$ which are eigenstates of
the operator $H$ of (\ref{H-sigmaL-hsigmatau})
with positive and negative eigenvalues.

Then, from (\ref{defgammahatprime}), we obtain
\begin{equation}
\langle j,m  | \hat\Gamma' | j,m \rangle
= \frac{h}{|h|} \left(
\begin{array}{ccc}
\mbox{\boldmath  $1$}_{2l+3}  &&    \\
  & X &    \\
 &&  \mbox{\boldmath  $1$}_{2l-1}        
\end{array}
\right) \ .
\end{equation}
For the block X of the $j=l$ sector,
we have to evaluate a sign of the coefficient 
$2 \cos \Bigl(2\theta(h)\Bigr)$ in
\beq
\left\{
U(h)\left(
\begin{array}{cc}
1  & 0   \\
0 & -1        
\end{array}
\right)U^{\dagger}(h) \ , \ 
\left(
\begin{array}{cc}
1  & 0   \\
0 & -1        
\end{array}
\right)
\right\} = 2 \cos \Bigl(2\theta(h)\Bigr) \, {\bf 1}_2 \ .
\label{UUc}
\eeq
Here we took $U(h)$ as
\beq
U(h) =
\left(
\begin{array}{cc}
\cos \bigl(\theta(h)\bigr)  & -\sin \bigl(\theta(h)\bigr)   \\
\sin \bigl(\theta(h)\bigr) & \cos \bigl(\theta(h)\bigr)       
\end{array}
\right) \ .
\label{UU}
\eeq

For $h=0$, the operator $H$ becomes $\sigma \cdot L + 1/2$,
and thus the basis $|\pm \rangle_H$
reduces to $|\pm \rangle_2$.
Then $U(h)$ becomes $U^{(12)}$, and we obtain
\beq
\cos \Bigl(2\theta(h=0)\Bigr)
=\frac{2}{n^2}-1 < 0 
\eeq
for $n \ge 2$.
Similarly, we can obtain
$
\cos \Bigl(2\theta(h=1)\Bigr)
=-1 < 0
$,
and
$
\cos \Bigl(2\theta(h=\infty)\Bigr)
=-1/n < 0
$.

We then study whether 
$\cos \Bigl(2\theta(h)\Bigr)=0$ 
takes place within the regions $(-2n+2)/3 <h <(2n+2)/3$.
$\cos \Bigl(2\theta(h)\Bigr)=0$ means 
$| \pm \rangle_H = (| + \rangle_1 \pm | - \rangle_1)/ \sqrt 2$.
This corresponds to the case where
 \beq
H (|+\rangle_{1} \pm |-\rangle_{1}  )
=e_\pm (|+\rangle_{1} \pm |-\rangle_{1}  )
\label{Hc}
\eeq
is satisfied at some value of $h$.
By using
\beqa
U^{(12)}\left( \begin{array}{cc}
l & \\  &-(l+1) 
\end{array} \right) U^{(21)}
&=&\frac{1}{(2l+1)^{2} }\left( 
 \begin{array}{cc}
l-4l(l+1)^{2} & 2\sqrt{l(l+1)}(2l+1)  \\ 
2\sqrt{l(l+1)}(2l+1) &  (4l^{2}-1)(l+1)
\end{array} \right)  \  , \nonumber \\
\\
U^{(13)}\left( \begin{array}{cc}
1 & \\  &-3 
\end{array} \right) U^{(31)}
&=&\frac{1}{2l+1 }\left( 
 \begin{array}{cc}
-2l-3 & -4\sqrt{l(l+1)}  \\ 
-4\sqrt{l(l+1)} &  -2l+1
\end{array} \right)   \ ,
\eeqa
(\ref{Hc}) is written as
\beqa
&&\left[ 
\frac{1}{(2l+1)^{2}} 
\left( 
 \begin{array}{cc}
l-4l(l+1)^{2} & 2\sqrt{l(l+1)}(2l+1)  \\ 
2\sqrt{l(l+1)}(2l+1) &  (4l^{2}-1)(l+1)
\end{array} \right)
 \right. \nonumber \\
&&~~\left.+\frac{h}{2}\frac{1}{2l+1 }\left( 
 \begin{array}{cc}
-2l-3 & -4\sqrt{l(l+1)}  \\ 
-4\sqrt{l(l+1)} &  -2l+1
\end{array} \right)
+\left( \frac{1}{2} -e_\pm \right) 
\left( \begin{array}{cc}
1 & \\ &1 
\end{array} \right) 
\right] \left( \begin{array}{c}
1 \\ \pm 1
\end{array}  \right) =0 \ .
\nonumber \\
\eeqa
Solving this equation,
we obtain 
\beqa
 h&=&-\frac{1}{2}(n^{2}-2) \ , \\
 e_\pm&=&\pm \frac{1}{2} \frac{1}{n} (n^{2}-1)^{\frac{3}{2}} 
 +\frac{1}{4n}(n^{3}-3n+2) \ .
\eeqa
For $n \ge 2$, $e_\pm$ take positive and negative values respectively,
which is consistent with the result in the previous subsection:
$H$ has always positive and negative eigenvalues in the $j=l$ sector,
since it does not have zero-modes for an arbitrary value of $h$.
Moreover, since $-(n^{2}-2)/2 < (-2n+2)/3$ for $n \ge 2$,
$\cos \Bigl(2\theta(h)\Bigr)=0$ does not take place
within the region  $(-2n+2)/3 <h < (2n+2)/3$.

Since the operator $H$ is a continuous function of $h$,
so is $\cos \Bigl(2\theta(h)\Bigr)$.
Thus $\cos \Bigl(2\theta(h)\Bigr)$ has always the same sign in
the region $(-2n+2)/3 <h < (2n+2)/3$.
Since $\cos \Bigl(2\theta(h)\Bigr)$ has a negative value 
at $h=0,1$, it always has negative values in
this region.
Therefore, we obtain our result (\ref{gammahatprime1-1-11}).

\end{document}